\documentclass[12pt,english]{revtex4}
\usepackage[T1]{fontenc}
\usepackage[latin1]{inputenc}
\usepackage{fancyhdr}
\usepackage{array}
\usepackage{float}
\usepackage{amsmath}
\usepackage{amssymb}

\makeatletter

\usepackage{graphicx}

\usepackage{babel}
\makeatother
\begin{document}

\title{Eyring equation and the second order kinetic law\\}

\author{\textit{L. Bonnet{\footnote{Corresponding author. Email: l.bonnet@ism.u-bordeaux1.fr}} and J.-C. Rayez}}

\address{Institut des Sciences Moléculaires, Université Bordeaux 1,
351 Cours de la Libération, 33405 Talence Cedex, France}

\begin{abstract}

Elementary gas-phase reactions of the bimolecular type A + B $\rightarrow$ Products are characterized 
by the second order kinetic law $-\frac{d[A]}{dt}=k[A][B]$,
where $[A]$ and $[B]$ are the concentrations of A and B species, $t$ is time and $k$ is the rate constant,
usually estimated by means of Eyring equation.
Here, we show that its standard derivation, as such, is not consistent with the second order law.
This contradiction is however removed by introducing a correlation between what we call \emph{potentially reactive pairs}.
A new derivation of Eyring equation is finally proposed on the basis of the previous findings.

\end{abstract}

\maketitle

\section{Introduction}

The kinetics of gas-phase elementary bimolecular reactions of the type A + B $\rightarrow$ Products is 
well known to be governed by the \emph{second order kinetic law}
\\
\begin{equation}
-\frac{d[A]}{dt}=k[A][B]
\label{1}
\end{equation}
\\
where $[A]$ and $[B]$ are the concentrations of A and B species, $t$ is time and $k$ is the \emph{rate constant}
\cite{Pratt, Atkins} (some details on the validity conditions of Eq. \eqref{1} are given later).

Since all the information concerning the kinetics is in $k$, predicting its value is a major goal
of theoretical chemists.

The most accurate way to perform such a prediction involves the simulation of electronic and nuclear motions.  
This is, however, a formidable task only feasible for a very limited number of processes \cite{Kendrick}.

On the other hand, \emph{Eyring equation} \cite{Eyring, Mahan}, 
the central result of \emph{transition state theory} (TST) 
\cite{Eyring, Mahan, Marcelin, Wigner, Polanyi, Wigner1, Horiuti, Eyring1, Laidler, Truhlar,Truhlar1}, 
allows the estimation of $k$ from very few data on the electronic structure of the system 
and with no need to simulate nuclear
motions. The computational cost is orders of magnitude lower than with the simulation. In addition to 
that, Eyring equation allows to pinpoint the key factors governing rate constants. Therefore, this equation has 
been very successful for several decades within the chemistry community \cite{Truhlar1}.

Quite surprisingly, its standard derivation appears not to be consistent with Eq. \eqref{1}, 
which happens to be annoying.
However, such a contradiction can be removed by introducing a \emph{correlation between potentially reactive pairs}.
The goal of the present note is to report these findings and propose on their basis a new derivation of Eyring equation.

\section{Theory}

\subsection{Molecular system}

Consider the elementary  bimolecular gas-phase reaction A + B $\rightarrow$ Products.
A and B are two molecules made of $I$ nuclei surrounded by electrons. 
The latter being much lighter than the formers, couplings between their respective motions can reasonably be ignored.
Electronic motions are then treated within the framework of quantum mechanics for fixed positions of the nuclei \cite{Anybook}. 
Different electronic states are possible. However, at the usual temperatures of kinetic experiments, 
the ground state is generally the only one to be populated, as we shall assume. In this state, nuclear motions are governed 
by the interaction potential $U$, sum of the electronic energy and the electrostatic energy between nuclei.
This potential is supposed to involve a barrier separating the reagent configurations from the product ones.

\subsection{Experimental conditions}

We consider a vessel of volume $V$ containing a gas made of about one mole of A, roughly the same quantity of B, 
and several moles of a given inert gas. 
Before time zero, the mixture is at very low temperature, in such a way that the probability for crossing the barrier is negligible. 
But let us assume that at time zero, we are able to instantaneously heat the gas in such a way that when two molecules A and B 
collide, their average amount of energy is now consistent with barrier crossing
(the temperature of the gas is supposed to be maintained after the heating).
$[A]$ and $[B]$ start then decreasing. Concomitantly, the newly formed products, two molecules C and D for instance, 
start diffusing in the gaseous medium to eventually collide with D and C molecules respectively, and reform the reagents 
A and B. However, this process takes some time, and we shall focus our attention on the initial period where it can be neglected.
In this period, it is well known that the kinetics is governed by Eq. \eqref{1} \cite{Pratt, Atkins}.

\subsection{Configuration space coordinates}

Expressing $U$ in terms of mass weighted Cartesian coordinates of the nuclei and diagonalizing the Hessian 
(matrix of second derivatives of $U$) at the barrier saddle point leads to a set of configuration space coordinates $q_i$, $i=\overline{1,3I}$, 
such that (i) the saddle point is the origin of the new frame, (ii) the second derivative of $U$ with respect to $q_1$ is negative 
and (iii) the same quantity for the remaining coordinates is positive. Motion is therefore unbounded along $q_1$ and bounded along the 
remaining $q_i$'s. $q_1$ is the steepest descent line around the saddle point, also called \emph{reaction coordinate}.
The barrier top is defined by $q_1 = 0$. Reagent (product) configurations correspond to negative (positive) values of $q_1$. 
$U^{\ddagger}$ is the value of $U$ at the saddle point.

\subsection{The three assumptions of transition state theory}

The three assumptions of TST are as follows \cite{Wigner1}: (i) we assume that nuclear motions can be described classically in the potential $U$;
(ii) thermalization is supposed to be much faster than reaction so that the reagents can be considered as thermalized 
at any instant; (iii) as soon as the system crosses the barrier top in the product direction, strongly repulsive forces tend to push 
the system away from the barrier top. It is thus reasonable to assume that A and B cannot be immediately reformed. If we define the 
transition state (TS) as the hyper surface of the phase space corresponding to $q_1 = 0$, 
the previous assumption can be restated as follows: the TS cannot be recrossed. Remember that 
we focus our attention on the initial period where the reagents cannot be reformed. Beyond this period, the TS can be recrossed
even if immediate recrossing is impossible.

\subsection{Phase space}

The momenta conjugate to the $q_i$'s are defined by $p_i = \partial L/\partial \dot{q}_i$, where
$L$ is the Lagrangian of the system (kinetic energy minus $U$) and $\dot{q}_i$ is the time derivative of $q_i$.

The dynamical state of the system is a point in the phase space, defined as the $6I$ dimensional space made of 
the ${q}_i$'s and the ${p}_i$'s.

The reagent (product) part of the phase space corresponds to the negative (positive) values of $q_1$.

\subsection{Phase space distribution of the dynamical states}

Assume that at a given instant $t$ posterior to the heating, the gas contains $N_A$ and $N_B$ molecules A and B, respectively. 
Since each A molecule can react with every one of the B molecules, there are 
\begin{equation}
N = N_A N_B
\label{2}
\end{equation}
\\
\emph{potentially reactive pairs}.
Since the gas is constantly thermalized, i.e., in canonical equilibrium, 
the dynamical states of the $N$ previous pairs are distributed in the phase space according to Boltzmann law. 
We shall call $\rho(\bold{q},\bold{p})$ the density of probability that a given pair has its state at 
the point $(\bold{q},\bold{p})$ of the reagent phase space, $\bold{q}$ gathering all the $q_i$'s and $\bold{p}$, all the $p_i$'s.
Stated differently, $\rho(\bold{q},\bold{p})$ is the probability of presence per unit volume of reagent phase space.

\subsection{Standard definition of the rate constant in TST}

The standard definition of the rate constant $k$ of the reaction is given by \cite{Eyring, Mahan}
\\
\begin{equation}
k=\frac{-\frac{d[N]}{dt}}{[A][B]}
\label{3}
\end{equation}
\\
with $[N]=N/V$, $[A]=N_A/V$ and $[B]=N_B/V$. This is exactly the expression given by Mahan \cite{Mahan}. Eyring uses
a different language, but his expression turns out to be equivalent \cite{Eyring}.

Using Eq. \eqref{2}, Eq. \eqref{3} can be rewritten as 
\\
\begin{equation}
\frac{k}{V}=\frac{-\frac{dN}{dt}}{N}.
\label{4}
\end{equation}

\subsection{Eyring equation}

$-\frac{dN}{dt}$ is the rate of disappearance of the potentially reactive pairs, i.e., 
the number of these pairs crossing the TS per unit time
(the minus sign is due to the fact that $N$ is decreasing). 
$-\frac{dN}{dt}/N$ is thus the proportion of these pairs crossing the TS per unit time,
also called reaction probability per unit time, or reaction probability flux.
Within the framework of the three assumptions of TST, this flux is given by the volume of 
reagent phase space crossing the TS per unit time, 
with each and everyone of its points weighted by the probability of presence per unit volume 
$\rho(\bold{q},\bold{p})$. This statement can be formalized as
\\
\begin{equation}
-\frac{1}{N}\frac{dN}{dt}=\int\;dS\;\dot{q}_1\;\Theta(\dot{q}_1)\;\rho(\bold{q},\bold{p}).
\label{5}
\end{equation}
\\
$dS$ is the element of the \emph{dividing surface} $q_1 = 0$,
another name for the TS,
$\dot{q}_1$ is the velocity along the reaction coordinate, i.e., perpendicular to the dividing surface, 
and $\Theta(\dot{q}_1)$ limits the integration to the 
points belonging to trajectories crossing the TS in the product direction. 

From Eqs. \eqref{4} and \eqref{5}, we arrive at
\\
\begin{equation}
k=V\int\;dS\;\dot{q}_1\;\Theta(\dot{q}_1)\;\rho(\bold{q},\bold{p}).
\label{6}
\end{equation}
\\
It can then be shown that Eq. \eqref{6} leads after some steps of algebra to Eyring equation \cite{Eyring, Mahan, Horiuti}
\\
\begin{equation}
k=\frac{k_B T}{h} \frac{Q^{\ddagger}}{Q_A Q_B}exp\Big(-\frac{U^{\ddagger}}{k_B T}\Big).
\label{7}
\end{equation}
\\
$h$ is Planck constant, $Q^{\ddagger}$ is the partition function per unit volume of the TS and $Q_A$ and $Q_B$ are
the same quantities for A and B.

\subsection{Kinetic law implied by the standard derivation}

Quite surprisingly, we have found no paper in which $N$ is replaced by $N_A N_B$ in Eq. \eqref{3}. When doing this substitution, 
Eq. \eqref{3} transforms to
\\
\begin{equation}
-N_B\frac{dN_A}{dt}-N_A\frac{dN_B}{dt}=kV[A][B].
\label{8}
\end{equation}
In addition to that,
\begin{equation}
\frac{dN_A}{dt}=\frac{dN_B}{dt},
\label{9}
\end{equation}
\\
for every time an A molecule disappears, a B molecule disappears too. Replacing $\frac{dN_B}{dt}$ by $\frac{dN_A}{dt}$ 
in Eq. \eqref{8} finally leads to 
\\
\begin{equation}
-\frac{d[A]}{dt}=\frac{k}{V}\frac{[A][B]}{[A]+[B]},
\label{10}
\end{equation}
\\
in strong disagreement with the expected Eq. \eqref{1}. 
The goal of the next subsection is to remove this contradiction.

\subsection{Introducing the correlation between potentially reactive pairs}

We still consider $N_A$ molecules A$_i$, $i=\overline{1,N_A}$, and $N_B$ molecules B$_j$, $j=\overline{1,N_B}$, during
the initial period where the reformation of the reagents has no time to take place and Eq. \eqref{1} is then valid. 

\begin{figure}[H]
\begin{center}\includegraphics[%
  clip,
  angle=0,
  origin=c]{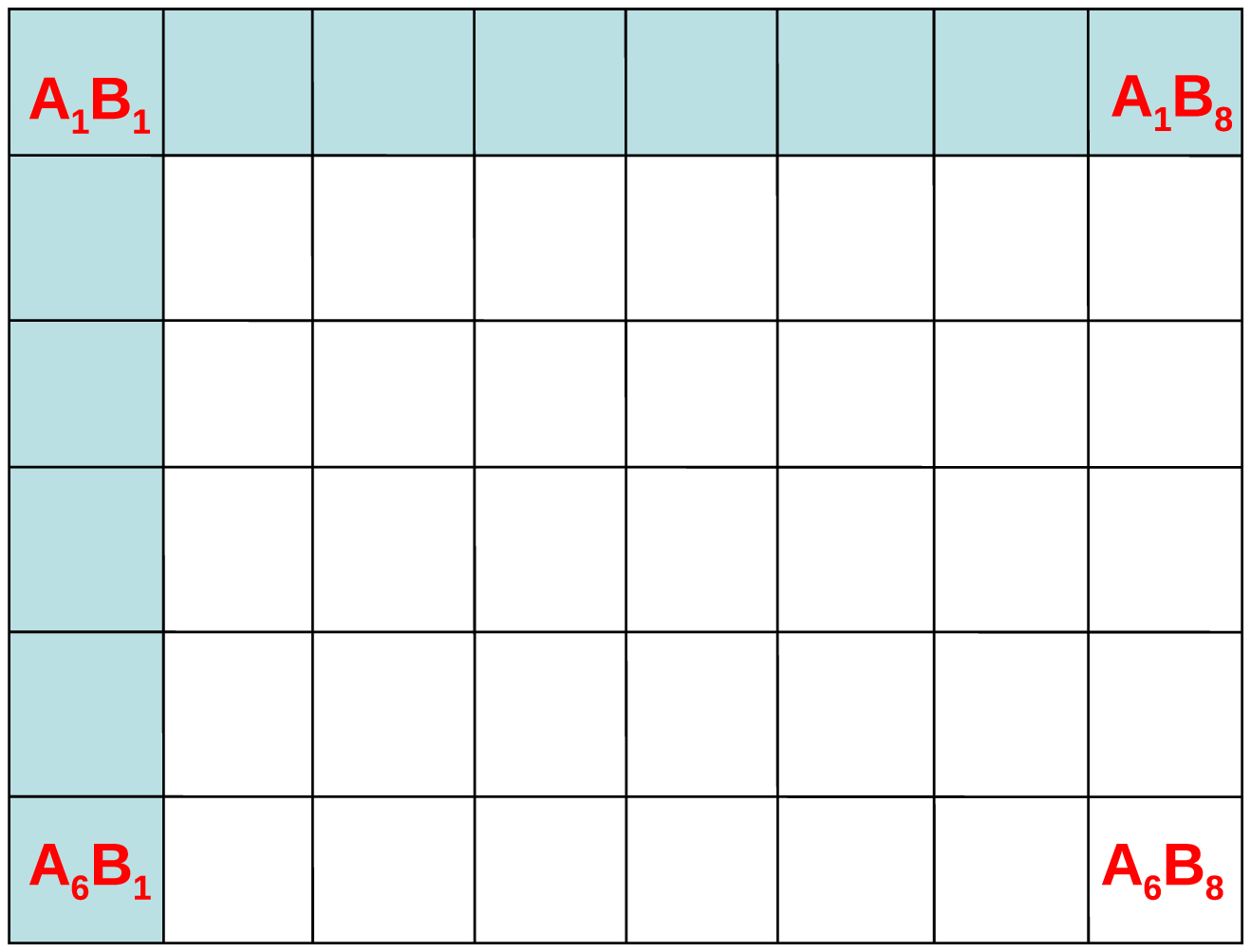}\end{center}

\caption{\label{fig:plot1-papier7}}
\end{figure}

Suppose that at a given instant $t$, A$_1$ reacts with B$_1$, i.e., the phase space point associated with the pair
A$_1$B$_1$ crosses the TS. At exactly the same time, the $(N_A+N_B-2)$ potential reactions between (i) A$_1$ and the $(N_B-1)$ 
molecules B$_j$, $j=\overline{2,N_B}$ 
and
(ii) B$_1$ and the $(N_A-1)$ 
molecules A$_j$, $j=\overline{2,N_A}$,
cease to be possible.
In other words, each time  
a potentially reactive pair leads to the products, $(N_A+N_B-2)$ analogous pairs become non reactive.
Overall, $(N_A+N_B-1)$ pairs must be removed from the potentially reactive pairs when one pair reacts.
The situation can be more easily understood with the help of the matricial drawing displayed in Fig. \ref{fig:plot1-papier7}.
$(N_A+N_B)$ being huge as compared to 1, $(N_A+N_B-1)$ will be approximated by $(N_A+N_B)$.

When such a correlation between potentially reactive pairs is ignored, we have seen that
within the framework of the three assumptions of TST, $N$ satisfies
\\
\begin{equation}
-\frac{dN}{dt}=\frac{k}{V} N,
\label{11}
\end{equation}
\\
with $k$ given by Eyring equation (see Eqs. \eqref{4}-\eqref{7}). 

If on the other hand, the previous correlation is taken into account, as it should be, 
the decrease of $N$ is $(N_A+N_B)$ times faster than previously, and Eq. \eqref{11} must be replaced by
\begin{equation}
-\frac{dN}{dt}= \frac{k}{V} N (N_A+N_B).
\label{12}
\end{equation}
\\
It is then an easy task to check from Eqs. \eqref{2} and \eqref{12}, that we
finally arrive at Eq. \eqref{1}.

The derivation of Eyring equation is now reconciled with the second order law.

\subsection{New derivation of Eyring equation}

We are now in a position to propose an alternative derivation of Eyring equation, which we believe to be
more satisfying than the standard derivation.

When ignoring the correlation between potentially reactive pairs, the three assumptions of TST lead
to
\begin{equation}
-\frac{dN}{dt}=N\int\;dS\;\dot{q}_1\;\Theta(\dot{q}_1)\;\rho(\bold{q},\bold{p})
\label{13}
\end{equation}
\\
(see Eq. \eqref{5}). On the other hand, the arguments of the previous subsection lead to
\\
\begin{equation}
-\frac{dN}{dt}=N\int\;dS\;\dot{q}_1\;\Theta(\dot{q}_1)\;\rho(\bold{q},\bold{p})(N_A+N_B)
\label{14}
\end{equation}
\\
when the correlation is taken into account.

From Eqs. \eqref{2} and \eqref{14}, we arrive at Eq. \eqref{1} with $k$ given by Eq. \eqref{6}, i.e., Eyring equation \eqref{7}.

In this derivation, there is no need to invoke the definition \eqref{3} of the rate constant, which to our 
mind, is not satisfying in view of Eq. \eqref{1}. Here, the second order law is integrated in the reasoning.

\section{Conclusion}

Rate constants of elementary gas-phase bimolecular reactions are usually estimated by using Eyring equation.

In this note, we show that its standard derivation, as such, is not consistent with the second order kinetic law
typical of elementary bimolecular processes. However, this contradiction turns out to be removed by introducing a correlation 
between potentially reactive pairs. 

A new derivation of Eyring equation consistent with the second order law is finally proposed on the basis of the previous findings.

\section*{Acknowledgments}

We thank Dr. M.-T. Rayez for helpful advices.

\newpage

\section*{Figures captions}

Fig. \ref{fig:plot1-papier7}: One may build an $N_A$ per $N_B$ matrix the elements of which are the potentially reactive pairs. 
Here, $N_A = 6$ and $N_B = 8$. For the sake of convenience, only 4 elements are shown.
When $A_{1}B_{1}$ reacts, the elements of the blue (or gray) line and column sharing $A_{1}B_{1}$ 
become non reactive. The total number of elements which must be removed from the potentially reactive pairs is thus 
$N_A+N_B-1 = 13$. \\

\newpage

\section*{Figures}

\newpage

\end{document}